%% file: main.tex
\documentclass[12pt]{article}

\usepackage[numbers]{natbib}
\usepackage{arxiv}

\usepackage{algorithm}
\usepackage{algpseudocode}

\usepackage{epic}
\usepackage{eepic}
\usepackage{paralist}
\usepackage{graphicx}
\usepackage{algorithm}
\usepackage{algpseudocode}
\usepackage{mathrsfs}
\usepackage{tikz}
\usepackage{xcolor,colortbl}
\usepackage{wrapfig}
\usepackage{amsmath}
\allowdisplaybreaks
\usepackage{amssymb}
\usepackage{mathtools}

\usepackage{graphicx}

\usepackage{doi}

\usepackage[toc,page]{appendix}

\input{preamble}

\usepackage{amsmath}
\usepackage{amsfonts}
\usepackage{amssymb}

\newcommand{\dg}[1]{\noindent\textup{\textsf{\color{red}[DG: #1]}}}

\title{Computing the Volume of a Restricted Independent Set Polytope Deterministically}

\iftrue
\author{ 
{
	\hspace{1mm}David Gamarnik}
 \thanks{Massachusetts Institute of Technology, Sloan School of Management, Operations Research Center, Institute for Data, Systems and Society. Support from NSF grants CISE-2233897 and DMS-2015517 is gratefully acknowledged}  
 \And
{
\hspace{1mm}
Devin Smedira}\thanks{Massachusetts Institute of Technology, Operations Research Center} 
}

\fi
\date{}

\renewcommand{\headeright}{}
\renewcommand{\undertitle}{}
\renewcommand{\shorttitle}{Volume of a Restricted Independent Set Polytope}

\hypersetup{
pdftitle={Temp},
pdfauthor={Devin ~Smedira, David~Gamarnik},
}

\usepackage{subfiles}
\begin{document}
\pagenumbering{roman}
\maketitle

\begin{abstract}
We construct a quasi-polynomial time deterministic approximation algorithm for computing the volume of an independent set polytope with  restrictions. Randomized polynomial time approximation algorithms for computing the volume of a convex body have been known now for several decades, but the corresponding deterministic counterparts are not available, and our algorithm is the first of this kind.
The class of polytopes for which our algorithm applies arises as linear programming relaxation of the independent set problem with the additional restriction that each variable takes value in the interval $[0,1-\alpha]$ for some $\alpha<1/2$. (We note that the 
$\alpha\ge 1/2$ case is trivial).

We use the correlation decay method for this problem applied to its appropriate and natural discretization. The method works provided $\alpha> 1/2-O(1/\Delta^2)$, 
where $\Delta$ is the maximum degree of the graph. When $\Delta=3$ (the sparsest non-trivial case), our method works provided $0.488<\alpha<0.5$. Interestingly, the interpolation method, which is based on analyzing complex roots of the associated partition functions, fails even in the trivial case when the underlying graph is a singleton.
\end{abstract}

\newpage
\pagenumbering{arabic}



\section{Introduction}
We consider the problem of designing deterministic polynomial time algorithm for the problem of computing the volume of a polytope. Namely, given
a polytope $\mathcal{P}$ described by a sequence of inequalities $\mathcal{P}=\{x: Ax\le b\}$,
where matrix $A$ and vector $b$ are given as problem input, the goal is to compute 
the volume (Lebeasgue measure) of $\mathcal{P}$
using deterministic algorithms with polynomial running time. 

It is known that solving this problem exactly
is $\#P$-hard even in very restricted 
cases~\cite{dyer1988complexity}, and thus it is unlikely that 
exact computation of the volume 
is possible with polynomial 
time algorithms.  The focus thus has 
shifted to designing approximation algorithms. In this
case the algorithms do exist, thanks to the celebrated
work by Dyer, Frieze and Kannan~\cite{DyerFriezeKannan}, who have
successfully constructed Markov Chain Monte Carlo (MCMC)
based algorithm which provides Fully Polynomial
Time Randomized Approximation Scheme (FPRAS),
when the problem input is any convex body, and 
not necessarily a polytope. 

FPRAS is nearly golden standard in the field of 
approximation algorithms, the only caveat being 
its reliance on randomization. This raises a question
whether a deterministic counterparts exist for the 
same problem. Unfortunately, in the case when
the input is a general convex body described with an
assistance of an appropriate oracle, deterministic
polynomial algorithms provably do not exist, thanks
to a clever argument by B{\'a}r{\'a}ny and F{\"u}redy~\cite{BaranyFuredi}. The special
case, though, when the input is explicitly given
as a polytope as opposed to oracle based description
is a totally different matter and here the existence
of deterministic approximation algorithms is not ruled 
out. This question is the central focus of this work.

The existence of deterministic counterparts to otherwise
randomized algorithms is a central question in 
theoretical computer science. It is widely believed
that any problem which admits polynomial time 
randomized algorithm (namely belongs to the either
of the classes RP or BPP) must admit a deterministic
counterpart (namely an algorithm in P)~\cite{trevisan2006pseudorandomness}. 
While proving this conjecture in full generality appears
out of reach, for many special cases this has been successfully
achieved, the problem of counting in graphs being one 
of them, which we turn to next. 

Recently we have witnessed a dramatic progress in designing
deterministic approximation algorithms 
for the problem of 
counting in graphs. Here the problem is as follows: given
a graph and given a combinatorial structure of a particular
type (independent set, matching, proper coloring, etc.) the 
goal is to compute approximately 
the number of such structures in the graph (exact computation
is \#P-hard again in most cases).
A generalization of this problem involves computing the 
object known as the partition function, a canonical object
arising in the field of statistical physics. 

For a while the only algorithms for solving this problem
was MCMC-based as thus relied on randomization.
In 2005 the correlation decay method was 
introduced simultaneously and independently by Bandyobapdhyay and Gamarnik~\cite{BandyopadhyayGamarnikCountingConference,BandyopadhyayGamarnikCounting} and Weitz~\cite{weitzCounting}, which led to deterministic approximation algorithms for counting in graphs. Loosely speaking
the correlation decay means,
that when the solution is sampled
from the associated Gibbs distribution (uniformly at random in particular), the marginal distribution associated with nodes at large distances are approximately product form. 
While the algorithm in~\cite{BandyopadhyayGamarnikCountingConference,BandyopadhyayGamarnikCounting} was restricted to graphs with growing girth, the algorithm 
in~\cite{weitzCounting}, did not have such restrictions and led for the first time to a Fully Polynomial Time Deterministic Approximation Scheme (FPTDAS) -- truly the strongest algorithm one could hope for for this problem. The algorithm was based on an extremely ingineous idea of  relating the problem of counting in graphs to the problem of counting in the associated
self-avoiding tree, appropriately defined. 
The exponent of the running time was improved recently to the best possible
by an important developments~\cite{anari2021spectral} based on spectral  independence. 

Later on another very 
powerful method for graph counting problems was introduced by Barvinok~\cite{barvinok2017combinatorics}. The method is based on viewing the partition function as
a complex variable polynomial and showing that the logarithm of this polynomial is well approximated by its low order Taylor terms provided the zeros of the polynomials are outside of a certain complex region containing $0$ and $1$. Remarkably, the power of the method evidently coincides with the correlation decay method, also leading to FPTDAS in the same regimes. A partial explanation for this was given in~\cite{gamarnik2022correlation} and~\cite{regts2023absence}.

Since the volume computation can be viewed as a idealized version of counting, it is only natural to attempt either  the correlation decay or the interpolation method to address this problem. We adopt the correlation decay method in this paper and, at the same time, show that the interpolation method runs into some technical barriers which we explain below.
We restrict our attention to polytopes which are linear programming relaxation of the independent set problem and introduce the set up next. We do conjecture though that the method extends far beyond this set up, although serious technical challenges do arise which we also discuss.

Consider a simple undirected graph $G$ on the node set $V$ and edge set $E$. An independent set is any subset of nodes $I\subset V$ which spans no edges. Namely,
for every $u,v\in E$ it is the case that $(u,v)\notin E$. Associate with every
node $u\in V$ a variable with range $[0,1]$
and with every edge $(u,v)$ a linear constraint $x_u+x_v\le 1$. This gives a rise to a polytope we denote by $\mathcal{P}_{\rm IS}(G)$. 
Computing the volume of this polytope
algorithmically is our focus.

If we consider a further restriction
$x_u\in \{0,1\}$ for every node and
count the number of solutions, this clearly
corresponds to the number of independent
sets in the graph $G$, which is the subject of~\cite{BandyopadhyayGamarnikCountingConference,BandyopadhyayGamarnikCounting,weitzCounting} and
many follow up papers. Indeed, Weitz' algorithm applies to this setting when the largest degree $\Delta$ of the graph $G$
does not exceed $5$. Importantly, in a dramatic development by Sly~\cite{sly2010computational}, it was shown that approximating the number
of independent sets in graphs when $\Delta$ is allowed to be at least  $6$ is NP-hard to compute. 

At the same time, the approximation algorithm of~\cite{DyerFriezeKannan} indicates that such hardness result does not hold for the volume computation problem. Thus, it is likely that the correlation decay holds for the model, appropriately defined. 
A partial progress in this direction was achieved in~\cite{gamarnik2019uniqueness}. There the independent set linear programming relaxation problem described above was considered when the underlying graph is a regular tree. It was established that indeed, consistently with the hypothesis above, the independent set polytope on regular trees exhibits correlation decay for \emph{all} degrees. Unfortunately, the method utilized in this paper is not adoptable to graphs with varying degree and new methods are needed. This is what we partially achieve in this paper.

We now describe our main result. We consider the problem of computing the volume of the the polytope $\mathcal{P}_{\rm IS}(G)$ with an additional assumption that each variable $x_v$ is restricted to be in the interval $[0,1-\alpha]$ for some $\alpha<1/2$. We denote this polytope by
$\mathcal{P}_{\rm IS}(G,\alpha)$.
We note that the case $\alpha\ge 1/2$ is trivial, as in this case the constraints $x_u+x_v\le 1$ become redundant and the volume of the polytope is $(1/2)^{|V|}$ trivially. The case $\alpha<1/2$ however is a different story and this is where we achieve a partial progress. We design a  deterministic approximation algorithm for this problem with the running time $|V|^{O(\log |V|/\epsilon)}$ for any fixed $\epsilon>0$. Namely, the value produced by the algorithm normalized by the correct volume is within the interval $(1-\epsilon,1+\epsilon)$. Thus our algorithm is quasi-polynomial time. We achieve this for the range of $\Delta$-dependent values of $\alpha$ where $\Delta$ is the maximum degree of the graph $G$. Specifically, when
$\Delta=3$, which corresponds to the sparsest graphs for which the problem is non-trivial, our algorithm applies when 
$\alpha>0.488$. Asymptotically, our algorithm works when $\alpha>1/2-O(1/\Delta^{2})$. Admittedly, the regime of applicability of our method is very narrow, in light of the fact that the problem trivializes when $\alpha=1/2$. Our analysis shows, however, that the problem is very subtle even in this very  narrow regime.

We now discuss our algorithm and the rationale for the restriction involving the parameter $\alpha$. Our algorithm is based on establishing the correlation decay property on the appropriately discretized version of the problem. Namely the problem of counting the number of solutions of the system of equations $x_u+x_v\le N, (u,v)\in E$ when $x_u$ are restricted to lie in the set $0,1,2,\ldots,N$ for large enough integer value $N$. An approach for proving the correlation decay property, specifically for non-binary models, is to establish a contraction of the 
functions  associated with computation of marginal values, appropriately defined. The
contraction in its own turn can be established by showing that  $\|\cdot\|_1$ norm of the gradient of this functions is  strictly smaller than $1$. 
This method of establishing the correlation decay property was introduced in~\cite{GamarnikKatz} and became a fairly canonical approach to similar problems later on. The earlier method introduced by Weitz~\cite{weitzCounting} is based on establishing a certain monotonicity property on the associated  self-avoiding tree, (a special case of the computation tree),  is evidently limited to  2-valued models, as opposed to the $N+1$-valued setting, which is the setting  in our case. 

Unfortunately, the $\|\cdot\|_1$-contraction \emph{does not} 
 hold for 
the model $\mathcal{P}_{\rm IS}(G)$, even 
in the special case when the underlying
graph is a regular tree, even though
the correlation decay property
does hold for this model
as per the main result 
in~\cite{gamarnik2019uniqueness}. This was verified 
by us numerically as we report in Section \ref{obstacles}. However,  introducing
the additional restriction of the form
$x_u\in [0,1-\alpha]$ does induce the
$\|\cdot\|_1$ contraction, and this constitutes the main technical result 
of our paper. The proof of the contraction
is somewhat involved, but fairly elementary.  

While we have established
the correlation decay property for this, admittedly a rather restricted case of polytopes, we do conjecture
that it holds in full generality. The existence
of a polynomial time approximation algorithms for
all convex bodies on the one
hand, the tight connection
between the problem of counting in graphs and correlation decay, on the other hand, strongly suggests the validity of this conjecture. Thus we pose the following conjecture: \emph{the Gibbs distribution associated with the Lebesgue measure of a polytope exhibits the correlation decay property}.
Resolving this conjecture even for the independent set polytope, or  polytopes associated with other combinatorial structures is a very interesting special case for this conjecture.

We close this section by noting that the alternative method of counting based
on the polynomial interpolation method does not appear to be effective for the problem of volume computation even in the trivial case when the underlying graph $G$ consists of a single node $u$ (with no loops). In this case, after the discretization the set of feasible solutions is trivially the entire range $0,1,2,\ldots,N$. The natural complex valued polynomial associated with this case is then 
$p(z)=\sum_{0\le k\le N}z^{k}=(1-z^{N+1})/(1-z)$. 
Unfortunately, this polynomial has the sequence of roots $z_i=e^{2\pi i k/(N+1)}, k=1,2,\ldots,N$ concentrating densely  around $1$.
This  prevents the construction of a  zero-free connected region containing $1$, which is the basis of the interpolation approach. We  conjecture, however, that this problem might be somewhat artificial and some clever modification of the interpolation method can be adopted which avoids this pitfall. What this alternative construction should look like, however, is not entirely clear to us, and we leave it as another very interesting open question.

We provide a brief overview of the remaining sections. Section \ref{definitions} will more formally define the problem outlined above and present the main result. Section \ref{preliminary} of this paper will present a discretization of the volume problem, and present some preliminary results. Section \ref{correlation decay} will present the proof of the core correlation decay property necessary to design the approximate counting algorithms. In Section \ref{algorithm}, we will present an algorithm and prove the first of our two core results below.
In Section \ref{volumes}, we will relate the discretization we introduce to volumes to prove the main result of this paper. Finally, Section \ref{obstacles} will discuss the barriers to extending our results to the polytopes $\P_{\text{IS}}(G)$. 

\section{Definitions and the Main Result} \label{definitions}

\subfile{sections/definitions}

\section{Preliminary Technical Results} \label{preliminary}

\subfile{sections/preliminaries}

\section{Proof of the Correlation Decay Property} \label{correlation decay}

\subfile{sections/correlation_decay}

\section{A Counting Algorithm} \label{algorithm}

\subfile{sections/algorithms}

\section{From Counting to Volumes} \label{volumes}

\subfile{sections/volumes}

\section{Barriers to the $\alpha = 0$ Case} \label{obstacles}

\subfile{sections/obstacles}

\bibliography{main}







\end{document}

%% file: preamble.tex
\usepackage{hyperref}
\usepackage{amsmath}
\usepackage{amsthm}

\renewcommand{\P}{\mathcal{P}}
\newcommand{\D}{\mathcal{D}}

\theoremstyle{definition}
\newtheorem{definition}{Definition}

\newtheorem{theorem}{Theorem}

\newtheorem{lemma}[theorem]{Lemma}

\newtheorem{proposition}[theorem]{Proposition}

\newcommand{\handout}[5]{
   \renewcommand{\thepage}{#1-\arabic{page}}
   \noindent
   \begin{center}
   \framebox{
      \vbox{
    \hbox to 5.78in { {\bf ORIE 6334 Bridging Continuous and Discrete Optimization} \hfill #2 }
       \vspace{4mm}
       \hbox to 5.78in { {\Large \hfill #5  \hfill} }
       \vspace{2mm}
       \hbox to 5.78in { {\it #3 \hfill #4} }
      }
   }
   \end{center}
   \vspace*{4mm}
}

\newcommand{\vol}{\operatorname{Vol}}

%% file: sections/definitions.tex
We repeat here the  definitions of the independent set polytope $\P_{\text{IS}}(G)$, as well as the restricted independent set polytope $\P_{\text{IS}}(G, \alpha)$ from the introduction for convenience.

Consider a simple undirected graph $G= (V,E)$ with $m=|V|$ vertices.
Let $\Delta$ denote the maximum degree of the graph. Namely, letting $d(v)=|\{u: (u,v)\in E\}|$ be the degree of the node $v$,
$\Delta=\max_{v\in V} d(v)$.
The independent set polytope $\P_{\text{IS}}(G)$ is defined as
\[\P_{\text{IS}}(G) = \{\x \in \R^m |  0\le x_v \leq 1, \forall v\in V, x_u + x_v \leq 1, \forall (u,v)\in E\}.
\]
Fix any Restriction Parameter $\alpha\in (0,1)$. The restricted independent set polytope $\P_{\text{IS}}(G, \alpha)$ is defined as
\[
\P_{\text{IS}}(G, \alpha) 
=
 \{\x \in \R^m |  0\le x_v \leq 1-\alpha, \forall v\in V, x_u + x_v \leq 1, \forall (u,v)\in E\}.
\]
We denote by $\vol(\P_{\text{IS}}(G, \alpha))$ the volume of this polytope. Specifically,
\begin{align*}
    \vol(\P_{\text{IS}}(G, \alpha))=
    \int 1\left(\x \in 
    \vol(\P_{\text{IS}}(G, \alpha))\right)d\lambda(\x),
\end{align*}
where $\lambda$ denotes the Lebesgue measure in $\R^m$. Our goal is computing $\vol(\P_{\text{IS}}(G, \alpha))$ algorithmically.

We define $\alpha_\Delta\in (0,1)$ as the smallest $\alpha$ satisfying 
\begin{align}\eql{bound-alpha}
 \frac{2 \Delta (1-\alpha)}{\alpha}\(\frac{(1-\alpha)^d}{\alpha^d} - 1 \) = 1.
\end{align}

We  now formally state the main  result of this work.
\begin{theorem}[Approximating the Volume of $\P_{\text{IS}}(G, \alpha)$]
\label{main result}
There exists an algorithm which for any graph
$G$, any $\alpha>\alpha_\Delta$,  and any $\epsilon>0$, produces a value $\tilde V$ satisfying
\begin{align*}
1-\epsilon
\le 
{\tilde V \over \vol(\P_{\text{IS}}(G, \alpha))}
\le 
1+\epsilon.
\end{align*}
The running time of the algorithm is 
$\left(\frac{m}{\epsilon}\right)^{O(\log(m/\epsilon))}$ time, where $m=|V|$ is the number of nodes in the graph. In particular, the algorithm is a Fully Quasi-Polynomial Time Deterministic Approximation Scheme (FQPTDAS). 
\end{theorem}
As mentioned in the introduction, we believe that this result can be improved to obtain a FPTDAS, namely removing the dependence on $m$ and $\epsilon$ in the exponent.
We note that several of the original approximation results based on the interpolation method were first of quasi-polynomial  time, and
later improved to be genuinely polynomial running time algorithms~\cite{patel2017deterministic}.

Next we introduce our main technical result, the correlation decay property on the associated computation tree, which is the basis of our algorithm and Theorem~\ref{main result}. Our first step is discretizing the problem and thus converting the problem of volume computation to the problem of counting.
Fix a large positive integer $N$, referred to as the Relaxation Parameter, and define the set 
\begin{align}\eql{PN}
\D_{\rm IS}(G,\alpha,N)
=
\{\x\in\Z_+^m: x_v\le (1-\alpha)N, \forall v \in V, x_u+x_v\le N, \forall (u,v)\in E\},
\end{align}
and the associated counting measure known as the partition function
\begin{align}\eql{ZG}
Z(G)=|\D_{\rm IS}(G,\alpha,N)|.
\end{align}
For linguistic convenience, we will refer the the vectors $\x \in \D_{\rm IS}(G,\alpha,N)$ as the Relaxed Independent Sets of the graph $G$.
Roughly speaking, as we will make more precise in Section~\ref{volumes}, the value of $Z(G)$ is the volume of $\P_{\rm IS}(G,\alpha)$ scaled by $N^m$. 
Our main focus is thus developing an approximation algorithm for estimating $Z(G)$. 

Due to the recursive nature of the algorithm, we will need to consider the set further restricted by a set of equalities of the form $x_v=n$. This type of constraint effectively eliminates one of the $n$ coordinates. 
A collection of such constraints will be generically denoted by $\beta$, and any time an additional constraint of the form $x_v=n$ is added, the collection is updated as $\beta'=\beta\cup (v\leftarrow n)$, which, with some abuse of notation, we write as $(\beta,v\leftarrow n)$.
Given a collection $\beta$ of such constraints we define $\D_{\rm IS}(G,\alpha,N,\beta)$ as the set of points in the set $\D_{\rm IS}(G,\alpha,N)$ satisfying the constraints in $\beta$. We will denote this set by $\D(G,\beta)$ and refer the its members as the Restricted Relaxed Independent Sets of $G$. We define by $Z(G,\beta)$ the cardinality of this set.  

We now introduce the Gibbs distribution associated with the counting measure $Z(G,\beta)$. This is just a uniform probability measure on the set $\D(G,\beta)$.
Namely, we assign probability $1/Z(G,\beta)$ to each point in $\D(G,\beta)$. For each node $v\in V$ and each integer $n\in [0,N]$, we denote by 
$x(G,v\leftarrow n,\beta)$ the probability that $x_v$ takes value $n$ under this measure. Note that
\begin{align*}
x(G,v\leftarrow n,\beta)={Z(G,(\beta,v\leftarrow n))\over Z(G,\beta)}.
\end{align*}
It is possible for both the numerator and the denominator of this expression to be 0. In this case, we define $x(G,v\leftarrow n,\beta)$ to be 0.

A standard approach for computing the partition function $Z(G)$ is to compute marginals $x(G,v\leftarrow n,\beta)$ up to inverse polynomial in $m$ multiplicative errors. This approach goes as follows. Fix any ordering of nodes $v_1,\ldots,v_n$ and fix any sequence of values $n_1,\ldots,n_m$ in $[0,(1-\alpha)N]$. Specifically we may take all of them equal to zero. Let $\beta_k$
be the sequence of constraints $v_j\leftarrow 0, 1\le j\le k$.
We have the following telescoping property
\begin{align*}
Z(G)&=
x^{-1}(G,v_1\leftarrow 0)Z(G,\beta_1) \\
&=
x^{-1}(G,v_1\leftarrow 0)x^{-1}(G,v_2\leftarrow 0, \beta_1)Z(G,\beta_2)\\
&=
\prod_{1\le k\le m}x^{-1}(G,v_k\leftarrow 0, \beta_{k-1}),
\end{align*}
under the convention that $\beta_0$ is an empty set of constraints. 
Suppose we algorithmically produce a sequence of estimations $y(G, v_k \la 0, \beta_{k-1}), 1\le k\le m$ of $x(G, v \la k, \beta_{k-1})$ up to multiplicative precision say $1/m^2$. Namely, 
\begin{align*}
1-1/m^2\le
{x(G, v_k \la 0, \beta_{k-1}) \over y(G, v_k \la 0, \beta_{k-1})}\le 1+1/m^2.
\end{align*}
Then $\hat Z(G) \triangleq \prod_k y^{-1}(G, v_k \la 0, \beta_{k-1})$ satisfies
\begin{align*}
(1-1/m^2)^m=1-O(1/m)\le 
{\hat Z(G)\over Z(G)} \le (1+1/m^2)^m
=1+O(1/m),
\end{align*}
and hence we obtain our goal of estimating $Z(G)$ up to multiplicative approximation better than any fixed constant $\epsilon$ for large $m$.
Thus our focus now is designing an algorithm for producing approximations
$y(G, v \la n,\beta)$ of $x(G, v \la n, \beta)$.

Our next step towards stating the correlation decay property is establishing a recursion structure exhibited by the marginal probabilities $x(G, v \la m, \beta)$. This recursive structure in full generality was introduced in~\cite{GamarnikKatz} and we adopt it here as well. 

We fix a graph $G=(V,E)$, node $v\in V$, and any set of constraints $\beta$. Let $v_1,\ldots,v_d$ be the neighbors of $v$, where $d=d(v)$ is the degree of the node. 
For any sequence of integers $n,n_1,\ldots,n_d\in [0,(1-\alpha)N]$, we denote $(n_1,\ldots,n_i)$ by $\bar n_i$ for short and  consider the associated marginal probabilities
\begin{align*}
x^i(\bar n_i)\triangleq
x\left(G\setminus v, v_i\leftarrow n_i, (\beta, \left(v_j\leftarrow n_j, 1\le j\le i-1\right)\right)).
\end{align*}
Namely, this is the probability that when nodes $v_1,\ldots,v_{i-1}$  are assigned values $n_1,\ldots,n_{i-1}$ respectively (thus updating the set of constraints from $\beta$ to $\beta,  v_j\leftarrow n_j, j\le i-1$), and $v$ is eliminated from $G$, the node $v_i$ receives value $n_i$.
The vector $(x^i(\bar n_i), 1\le i\le d)$ of marginals is denoted by $\x$.
For every $k$, we let $S_k=[0,\min\left(N-k,(1-\alpha)N\right)]$. 
Define the sequence 
$\left(f_n, n\in [0,(1-\alpha)N]\right)$  as follows:
\begin{align}\eql{f_n}
    f_n(\x) = \frac{\sum\limits_{n_1, \ldots n_d \in S_n} x^1(\bar{n}_1) \ldots x^d(\bar{n}_d)}
{\sum\limits_{m=0}^{(1-\alpha)N} \sum\limits_{n_1, \ldots n_d \in S_m} x^1(\bar{n}_1) \ldots x^d(\bar{n}_d)} .
\end{align}
The recursive structure is stated in the following proposition which will be proved in the next section.

\begin{proposition}
\label{recursion}
The following relation holds for every 
$n\in [0,(1-\alpha)N]$:
\begin{align*}
x(G,v\leftarrow n, \beta)=f_n(\x).
\end{align*}
\end{proposition}

Ultimately, we will use the recursion (\ref{eq:f_n}), to compute estimations $\y$ of the true marginal probabilities $\x$ recursively. We will introduce the notion of a Feasible Marginal Distribution for this problem setup in Section \ref{initial observations}, and show that both the true marginals $\x$ and the recursive estimates $\y$ satisfy the definition. 
Our main correlation decay result states that provided two vectors $\x$ and $\y$ are Feasible Marginal Distributions and $\alpha$ satisfies (\ref{eq:bound-alpha}), the following contraction property
holds for every $n\in [0,(1-\alpha)N]$ and for some $0<c=c(\alpha)<1$,
\begin{align*}
|f_n(\y)-f_n(\x)|\le c \|\y-\x\|_\infty.
\end{align*}
Namely, every iteration of the recursion encoded in $f_n$ decreases the estimation error geometrically. Provided the number of iterations is of the order $\Theta(\log n)$, we will obtain the required upper $1/n^{\Omega(1)}$ on the estimation error. This is the correlation decay on the computation tree property which
is the main technical underpinning of our work.

%% file: sections/preliminaries.tex
This section will be primarily focused on providing a complete proof of Proposition \ref{recursion}. We will then provide a definition of a Feasible Marginal Distribution and establish a few related properties.  

\subsection{Proof of Proposition \ref{recursion}}

We will now show how to write an expression for $x(G, v \la n, \beta)$ in terms of the variables $x^i(\bar{n}_i)$, which represent the probability distribution over the assignments of adjacent vertices in smaller graphs. We will again assume that the vertex $v$ has $d \leq \Delta$ neighbors $v_1, \ldots v_d$. We start this process by proving the next two Lemmas.

\begin{lemma}[Recursive Deconstruction]
\label{recursive}
\begin{align*}
x(G, v \la n, \beta) =& x(G, v \la 0, \beta) \sum\limits_{n_1, \ldots n_d \in S_n} x(\bar{n}_1) \ldots x(\bar{n}_d)
\end{align*}
\end{lemma}
\begin{proof}

Notice that 
\[Z(G, (\beta, v \la 0)) = Z(G \setminus v, \beta),\]
and more generally
\[Z(G, (\beta, v \la n)) = \sum\limits_{(n_1, ... n_d) \in S_n^d} Z(G \setminus v, (\beta, (v_i \la n_i \forall i))).\]
Using this, we have
\begin{align*}
&x(G, v \la n, \beta)\\
&= \frac{Z(G, (\beta, v \la n))}{Z(G,\beta)}\\
&= \frac{Z(G, (\beta, v \la 0)) Z(G, (\beta, v \la n))}{Z(G, \beta) Z(G \setminus v, \beta)}\\
&=\frac{x(G, v \la 0, \beta)}{Z(G \setminus v, \beta)} Z(G, (\beta, v \la n))\\
&= \frac{x(G, v \la 0, \beta)}{Z(G \setminus v, \beta)} \sum\limits_{(n_1, ... n_d) \in S_n^d} Z(G \setminus v, (\beta, (v_i \la n_i \forall i)))\\
&= \frac{x(G, v \la 0, \beta)}{Z(G \setminus v, \beta)} \sum\limits_{n_1 \in S_n} Z(G\setminus v, (\beta, v_1 \la n_1)) \sum\limits_{(n_2, ... n_d) \in S_n^{d-1}} 
\frac{Z(G \setminus v, (\beta, (v_i \la n_i \forall i)))}{Z(G\setminus v, (\beta, v_1 \la n_1))}\\
&= x(G, v \la 0, \beta) \sum\limits_{n_1 \in S_n} {Z(G\setminus v, (\beta, v_1 \la n_1)) \over Z(G \setminus v, \beta)} \sum\limits_{(n_2, ... n_d) \in S_n^{d-1}} 
\frac{Z(G \setminus v, (\beta, (v_i \la n_i \forall i)))}{Z(G\setminus v, (\beta, v_1 \la n_1))}\\
&= x(G, v \la 0, \beta) \sum\limits_{n_1 \in S_n} x(G\setminus v, v_1 \la n_1, \beta) \sum\limits_{(n_2, ... n_d) \in S_n^{d-1}} \frac{Z(G \setminus v, (\beta, (v_i \la n_i \forall i)))}{Z(G\setminus v, (\beta, v_1 \la n_1))}\\
&= x(G, v \la 0, \beta) \sum\limits_{n_1 \in S_n} x(\bar{n}_1) \sum\limits_{(n_2, ... n_d) \in S_n^{d-1}} \frac{Z(G \setminus v, (\beta, (v_i \la n_i \forall i)))}{Z(G\setminus v, (\beta, v_1 \la n_1))}\\
&= x(G, v \la 0, \beta) \sum\limits_{n_1 \in S_n} x(\bar{n}_1) \ldots \sum\limits_{n_d \in S_n} x(\bar{n}_d) \\
&= x(G, v \la 0, \beta) \sum\limits_{n_1 , \ldots n_d\in S_n} x(\bar{n}_1) \ldots x(\bar{n}_d),
\end{align*}

where the second to last line is derived by repeating the same logic for each neighbor $v_i$ with a, appropriately modified value of $\beta$. 
\end{proof}

\begin{lemma}[Normalizing Term]
\label{normalizing}
\[x(G, v \la 0, \beta) = \frac{1}{\sum\limits_{m=0}^{(1-\alpha)N} \sum\limits_{n_1 , \ldots n_d\in S_m} x(\bar{n}_1) \ldots x(\bar{n}_d)}.\]
\end{lemma}
\proof By construction, $\sum\limits_{m=0}^{(1-\alpha)N} x(G, v \leftarrow m, \beta) = 1$, since $((x(G, v\la m, \beta))$ is a probability vector for $m \in [0,(1-\alpha)N]$. Thus, by Lemma \ref{recursive},
\begin{align*}
1 =  \sum\limits_{m=0}^{(1-\alpha)N} x(G, v \leftarrow m, \beta)
= x(G, v \la 0, \beta) \sum\limits_{m=0}^{(1-\alpha)N} \sum\limits_{n_1 , \ldots n_d\in S_m} x(\bar{n}_1) \ldots x(\bar{n}_d).
\end{align*}
Rearranging terms, we have
\[x(G, v \la 0, \beta) = \frac{1}{ \sum\limits_{m=0}^{(1-\alpha)N} \sum\limits_{n_1 , \ldots n_d\in S_m} x(\bar{n}_1) \ldots x(\bar{n}_d)}.\]
\qed\\

Combining Lemmas \ref{recursive} and \ref{normalizing}, we immediately obtain Proposition \ref{recursion}.

\subsection{Bounds on Marginal Probabilities}\label{initial observations}

Section \ref{correlation decay} will be dedicated to establishing a bound on the 1-norm of the function $f_n$ from equation \ref{eq:f_n}. The bound we compute will need to hold not just on the true probability distribution $\x$, but also on any linear interpolation between the true distribution and our probability estimates. To do this, we identify the primary properties we need for the contraction to hold. Thus, we introduce the following notion of a Feasible Marginal Distribution, and show that our bound will hold on all such distributions.

\begin{definition}[Feasible Marginal Distribution]\label{feasible marginal distribution}
Fix some $\w = \{w_k > 0, 0 \leq k \leq (1- \alpha) N\}$. We say that $\w$ is a \textit{Feasible Probability Distribution} if
\begin{alignat*}{3}
\sum\limits_{k = 0}^{(1-\alpha)N} w_k &= 1, &\text{   }& \eql{1 sum} \\
w_k &\geq w_{k+1}, &&\eql{non-inc}\\
w_{k} &= w_{k'}  &\text{ }& \forall k,k' \leq \alpha N.  \eql{alpha N}
\end{alignat*}
Further, we say $\y = \{y^j(\bar{n}_j)| j \in \{1, \ldots d\}, \bar{n}_j \in [0,(1-\alpha)N]^j\}$ is a \textit{Feasible Marginal Distribution} if for every $j \in \{1, \ldots d\}$ and $\bar{n}_{j-1} \in [0, (1-\alpha)N]^{j-1}$ where $(\beta, v_i \la n_i, i < j)$ forms a feasible set of constraints, the values $y_i = y^j(\bar{n}_{j-1}, i)$ for $i \in [0, (1-\alpha)N]$ form a Feasible Probability Distribution and for every $\bar{n}_j, \bar{m}_j \in [0, \alpha N]$, 
\[y^j(\bar{n}_j) = y^j(\bar{m}_j). \eql{marginals}\]

\end{definition}


We now show that $\x$ as defined above is a Feasible Marginal Distribution. Firstly, observe from the problem's construction that for any fixed $j, n_1, \ldots n_{j-1}$, the set $\{x^j(\bar{n}_{j-1}, n_j)| n_j \in [0,(1-\alpha)N] \}$ forms a probability distribution when $(\beta, v_i \la n_i, i < j)$ forms a feasible set of constraints, meaning equation \ref{eq:1 sum} is satisfied. Next, notice that $Z(G, (\beta, v \la n)) \geq Z(G, (\beta, v \la n+1))$, since for every relaxed independent set where $v$ is assigned $n+1$ there is a relaxed in dependent that is identical except $v$ is assigned $n$. Thus, we have that for every fixed $\bar{n}_{j-1}$, $x(\bar{n}_{j-1},n_j) \geq x(\bar{n}_{j-1},n_j + 1)$, so $\x$ satisfies equation \ref{eq:non-inc}. Finally, we have that $Z(G, (\beta, v \la n)) = Z(G, (\beta, v \la n'))$ for every $0 \leq n,n' \leq \alpha N$, since assigning any vertex a value in $[0, \alpha N]$ cannot violate any constraints. More generally, if every constraint in $\beta$ and $\beta'$ assigns a value of at most $\alpha N$, then $Z(G, \beta) = Z(G, \beta')$. This implies $\x$ must satisfy equations \ref{eq:alpha N} and \ref{eq:marginals}. Therefore, $\x$ is a Feasible Marginal Distribution. 

\begin{lemma}[Closure of Feasible Marginal Distributions under $f_n$]\label{closure f}
Let $\y$ be a Feasible Marginal Distribution, and let $y_n' = f_n(\y)$ for $n \in [0, (1-\alpha)N]$. Then, the set of $y_n'$ form a Feasible Probability Distribution.
\end{lemma}

\begin{proof}
Notice that $S_n = S_{n'}$ for any $n,n' \leq \alpha N$ by construction. By equation \ref{eq:f_n}, we can clearly see this implies $f_n(\y) = f_{n'}(\y)$ when $n,n' \leq \alpha N$, so the values of $y_n'$ satisfy equation \ref{eq:alpha N}. Equation \ref{eq:1 sum} is satisfied because 
\[\sum\limits_{n=0}^{(1-\alpha)N} f_n(\y) = \frac{\sum\limits_{n=0}^{(1-\alpha)N} \sum\limits_{n_1, \ldots n_d \in S_n} x^1(\bar{n}_1) \ldots x^d(\bar{n}_d)}
{\sum\limits_{m=0}^{(1-\alpha)N} \sum\limits_{n_1, \ldots n_d \in S_m} x^1(\bar{n}_1) \ldots x^d(\bar{n}_d)} = 1.\]
Finally, equation \ref{eq:non-inc} is clearly satisfied because $S_n$ is non-increasing in $n$ and each term in the numerator sum is non-negative.
\end{proof}

\begin{lemma}[Linear Interpolations of Feasible Marginal Distributions]\label{linear interpolations}
Let $\y$ and $\z$ be two Feasible Marginal Distributions. Then, for any $w \in [0,1]$, $w \y + (1 - w) \z$ is a Feasible Marginal Distribution, where the sum is taken pointwise.
\end{lemma}

The proof of Lemma \ref{linear interpolations} follows directly from the linearity of the four equations \ref{eq:1 sum}, \ref{eq:non-inc}, \ref{eq:alpha N}, and \ref{eq:marginals}. These results will allow us to construct the contraction in sufficient generality. We conclude by proving two Lemmas about Feasible Probability Distributions.

\begin{lemma}
\label{outcome}
For any Feasible Probability Distribution $\y$, we have
\[y_i \leq \frac{1}{\alpha N}.\]
\end{lemma}
\begin{proof}
By equation \ref{eq:alpha N}, for any $i,j \in [0, \alpha N]$, the value of $y_i = y_j$. There are $\alpha N$ such values, so each one must be no greater than $\frac{1}{\alpha N}$. Additionally, for any $i$, $y_i \leq y_0 \leq \frac{1}{\alpha N}$ by equation \ref{eq:non-inc}, finishing the proof.
\end{proof}

\begin{lemma}
\label{probability}
For any Feasible Probability Distribution $\y$, we have
\[ \frac{\alpha}{1-\alpha} \le \sum_{i=0}^{\alpha N} y_i \le 1. \]
\end{lemma}
\begin{proof}
The upper bound holds directly from equation \ref{eq:1 sum}.
To prove the lower bound,  first observe equation \ref{eq:1 sum} implies $\sum\limits_{i = 0}^{(1-\alpha)N} y_i = 1$. If $\sum\limits_{i = 0}^{\alpha N} y_i < \frac{\alpha}{1 - \alpha}$, then it would have to be the case that  $\sum\limits_{i = \alpha N + 1}^{(1 - \alpha) N} y_i > \frac{1 - 2\alpha}{1-\alpha}$. This would imply the last $1 -2\alpha$ terms in the sequence had a larger average value than the first $\alpha$ terms. Equation \ref{eq:non-inc} implies this sequence must be decreasing, which is a contradiction. Thus, the bound must hold. 

\end{proof}

%% file: sections/correlation_decay.tex

In this section, we establish the key correlation decay property necessary to design the approximation algorithm. The primary result, occurring at the end of the section, will be a bound on $\|f_n(\y)\|_1$ defined in equation \ref{eq:f_n} with respect to a vertex $v$ with $d$ neighbors for any Feasible Marginal Distribution $\y$. The bound will be in terms of $\alpha$ and $d$, and will be bounded away from 1 as $\alpha$ approaches $1/2$ for any fixed $d$. 
\subsection{Expressions for Partial Derivatives}

We now introduce the following two expressions to represent the numerator and denominator of equation \ref{eq:f_n}:
\begin{align*}
U_n &= \sum\limits_{n_1, \ldots n_d \in S_n} y^1(\bar{n}_1) \ldots  y^d(\bar{n}_d),\\
L &= \sum\limits_{m=0}^{(1-\alpha)N} \sum\limits_{n_1, \ldots n_d \in S_m} y^1(\bar{n}_1) y^d(\bar{n}_d) = \sum\limits_{m=0}^{(1-\alpha)N} U_m.
\end{align*}
In particular, we have that $f_n(\y) = \frac{U_n}{L}$. By the chain rule, we can see
\[\frac{\partial f_n(\y)}{\partial y^j(\bar{n}_j)} = \frac{L \Udev{j} - U_n \Ldev{j}}{L^2} \eql{derivative}.\]
Let 
\begin{align*}
B_j(\bar{n}_{j-1}) &= y^1(\bar{n}_1) \ldots y^{j-1}(\bar{n}_{j-1}),\\
A_j(n, \bar{n}_j) &= \sum\limits_{n_{j+1}, \ldots n_d \in S_n} y^{j+1}(\bar{n}_{j+1}) \ldots y^{d}(\bar{n}_{d}).
\end{align*}
Thus, for any $j \in [1,d]$, we have
\begin{align*}
U_n &= \sum\limits_{n_1, \ldots n_{j-1} \in S_n} B_j(\bar{n}_{j-1}) \sum\limits_{n_j \in S_n} y^j(\bar{n}_j) A_j(n,\bar{n}_j) \eql{Un},\\
L &= \sum\limits_{m=0}^{(1-\alpha)N} \sum\limits_{n_1, \ldots n_{j-1} \in S_m} B_j(\bar{n}_{j-1}) \sum\limits_{n_j \in S_m } y^j(\bar{n}_j) A_j(m,\bar{n}_j) \eql{L}.
\end{align*}
Fix $j \in [1,d]$ and some $\bar{n}_j \in [0, (1-\alpha)N]^j$. Letting $I(A)$ be the indicator variable for the event $A$, we obtain
\begin{align*}
\frac{\partial U_n}{\partial x^j(\bar{n}_j)} &=  I(\bar{n}_j \in S_n^d) B_j(\bar{n}_{j-1}) A_j(n, \bar{n}_j),  \eql{DUn}\\
\frac{\partial L}{\partial x^j(\bar{n}_j)} &= \sum\limits_{m=0}^{(1-\alpha)N} I(\bar{n}_j \in S_m^d) B_j(\bar{n}_{j-1}) A_j(m, \bar{n}_j). \eql{DL}
\end{align*}

\begin{lemma}
\label{AB Bound}
For any $j \in [1,d]$, $n \in [0,(1-\alpha)N]$, and $n_1, \ldots n_{j} \in [0,(1-\alpha)N]$, we have
\begin{align*}
A_j(n, \bar{n}_{j}) &\leq 1,\\
B_j(\bar{n}_{j-1}) &\leq \frac{1}{(\alpha N)^{j-1}}.
\end{align*}
\end{lemma}
\begin{proof}
Recall that for any $\bar{n}_{d-1}$, $y_d(\bar{n}_d)$ is a probability distribution as $n_d$ varies. Thus, 
\[\sum\limits_{n_d \in S_n} y_d(\bar{n}_d) \leq  \sum\limits_{n_d=0}^{(1-\alpha)N} y_d(\bar{n}_d) = 1.\]
Iterating this process for vertices $v_d, v_{d-1}, \ldots v_{j+1}$ leads to the bound on $A_j$. The bound on $B_j$ follows from Lemma \ref{outcome}.
\end{proof}

\begin{lemma}
\label{AB Same}
For any $j \in [1,d]$, $n \in [0,(1-\alpha)N]$, $n_1, \ldots n_{j} \in [0,\alpha N]$, and $m_1, \ldots m_{j} \in [0,\alpha N]$ we have that
\begin{align*}
A_j(n, \bar{n}_j) &= A_j(n, \bar{m}_j)\\
B_j(\bar{n}_{j-1}) &= B_j(\bar{m}_{j-1})
\end{align*}
\end{lemma}
\begin{proof}
By definition, we have $y^{j'}(\bar{n}_{j'}) = y^{j'}(\bar{m}_{j'})$ for any $j' \le j$. Repeatedly substituting these equalities into the definitions of $A_j$ and $B_j$ produces the desired result.
\end{proof}

We now move to bounding our numerator and denominator functions, $U_n$ and $L$, directly.

\begin{lemma}[Gradient Denominator]
\label{denominator}
\[L \geq \alpha N\]
\end{lemma}
\begin{proof}
Fix any $m \leq \alpha N$. We will have that $S_m = [0, \min(N-m, (1-\alpha)N] = [0,(1-\alpha)N]$. For any $n_1, \ldots n_{j-1}$, $\sum\limits_{n_j = 0}^{\alpha N} y^j(\bar{n}_j) = 1$ since $\y$ is a Feasible Marginal Distribution. Thus, 
\begin{align*}
\sum\limits_{n_1, \ldots n_d \in S_m} y^1(\bar{n}_1) \ldots y^d(\bar{n}_d)
= \sum\limits_{n_1, \ldots n_d=0}^{(1-\alpha)N} y^1(\bar{n}_1) \ldots y^d(\bar{n}_d)
= 1.
\end{align*}
We then obtain
\begin{align*}
L = \sum\limits_{m=0}^{(1-\alpha)N} \sum\limits_{n_1, \ldots n_d \in S_m} y^1(\bar{n}_1) \ldots y^d(\bar{n}_d)
\geq \sum\limits_{m=0}^{\alpha N} \sum\limits_{n_1, \ldots n_d \in S_m} y^1(\bar{n}_1) \ldots y^d(\bar{n}_d)
= \sum\limits_{m=0}^{\alpha N} 1 
= \alpha N.
\end{align*}
\end{proof}

\begin{lemma}[Upper bounds]
\label{upper}
For any $j \in [1,d]$, $n \in [0,(1-\alpha)N]$, and $n_1, \ldots n_{j-1} \in [0,(1-\alpha)N]$, we have
\begin{align*}
U_n &\leq 1, \\
L &\leq (1-\alpha)N, \\
\Udev{j} &\leq \frac{1}{(\alpha N)^{j-1}},  \\ \Ldev{j} &\leq \frac{(1-\alpha)N}{(\alpha N)^{j-1}}.
\end{align*}
\end{lemma}
\begin{proof}
We will show the bounds on $U_n$ and $\Udev{j}$. These will immediately imply the bounds on $L$ and $\Ldev{j}$ by definition. Writing $U_n$ in terms of $A_1$ in equation \ref{eq:Un}, we see that
\begin{align*}
U_n = \sum\limits_{n_1 \in S_n} y^j(\bar{n}_1) A_1(n, \bar{n}_1)
\leq \sum\limits_{n_1 \in S_n} y^j(\bar{n}_1) 
\leq 1,
\end{align*}
where the first inequality follows from Lemma \ref{AB Bound} and the second follows from equation \ref{eq:1 sum}. Similarly, we can write $\Udev{j}$ in terms of $A_j$ and $B_j$ as in equation \ref{eq:DUn} to see that
\[\Udev{j} = I(\bar{n}_j \in S_n^d) B_j(\bar{n}_{j-1}) A_j(n, \bar{n}_j) \leq  B_j(\bar{n}_{j-1}) A_j(n, \bar{n}_j) \leq \frac{1}{(\alpha N)^j},\]
where the last inequality follows from Lemma \ref{AB Bound}.

\end{proof}

\subsection{Upper Bounds on Partial Derivatives}
We are now obtain bounds on the value of the partial derivatives of $f(\y)$, in order to ultimately bound $\|\nabla f(\y)\|_1 $. We will do this by using two different bounds on $\frac{\partial f(\y)}{\partial y^j(\bar{n}_j)}$, depending on the value of $\bar{n}_j$. We present the looser bound first, which applies to all partial derivatives.
\begin{theorem}[Loose Partial Bound]
\label{loose}
For any $j \in [1,d]$ and $\bar{n}_j \in [0,(1-\alpha)N]^j$,
\[\left | \frac{\partial f(\y)}{\partial y^j(\bar{n}_j)} \right | \leq \frac{(1-\alpha) N}{(\alpha N)^{j+1}}\]
\end{theorem}
\begin{proof}
As we saw in equation \ref{eq:derivative}, we have 
\begin{align*}
\left |\frac{\partial f(\y)}{\partial y^j(\bar{n}_j)} \right | &= \left |\frac{L \Udev{j} - U_n \Ldev{j}}{L^2} \right |\\
&\leq \frac{\max(L \Udev{j}, U_n \Ldev{j}}{L^2}\\
&\leqa \frac{\max(L \Udev{j}, U_n \Ldev{j})}{(\alpha N)^2}\\
&\leqb \frac{(1-\alpha) N}{(\alpha N)^{j+1}},
\end{align*}
where inequality a follows from Lemma \ref{denominator} and  inequality b follows from Lemma \ref{upper}.
\end{proof}
As we will see in the next section, this bound alone will not be sufficiently tight. Next, we introduce a tighter bound on the partial derivative for cases where $\bar{n}_j \in [0,\alpha N]^j$. 

\begin{theorem}[Tighter Partial Bound] 
\label{tighter}
For any $j \in [1,d]$ and $\bar{n}_j \in [0,\alpha N]^j$,
\[\left|\frac{\partial f(\y)}{\partial y_j(\bar{n}_j)} \right | \leq \frac{j(1-2\alpha)(1-\alpha)^j}{\alpha^{2j+1} N^{j}}.\]
\end{theorem}

\begin{proof}

\begin{align*}
&\left|\frac{\partial f(\y)}{\partial y_j(\bar{n}_j)} \right |\\
&= \frac{|L \Udev{j} - U_n \Ldev{j}|}{L^2}\\
&\leqa \frac{1}{(\alpha N)^2} |L \Udev{j} - U_n \Ldev{j}|\\
&\eqb \frac{1}{(\alpha N)^2} \left |\sum\limits_{m=0}^{(1-\alpha)N}\left [ 
\sum\limits_{\bar{m}_j \in S_m^j} \Udev{j} B_j(\bar{m}_{j-1})y^j(\bar{m}_j) A_j(m, \bar{m}_j) - U_n I(\bar{n}_j \in S_m^j) B_j(\bar{n}_{j-1}) A_j(m, \bar{n}_j)
\right ] \right |\\
&\eqc \frac{1}{(\alpha N)^2} \Bigg |\sum\limits_{m=0}^{(1-\alpha)N}\Bigg [ 
\sum\limits_{\bar{m}_j \in S_m^j} I(\bar{n}_j \in S_n^j) B_j(\bar{n}_{j-1}) A_j(n, \bar{n}_j) B_j(\bar{m}_{j-1})y^j(\bar{m}_j) A_j(m, \bar{m}_j) \\
&- \sum\limits_{\bar{p}_j \in S_n^j} B_j(\bar{p}_{j-1}) y^j(\bar{p}_j) A_j(n, \bar{p}_j) I(\bar{n}_j \in S_m^j) B_j(\bar{n}_{j-1}) A_j(m, \bar{n}_j)
\Bigg ] \Bigg | \\
&\eqd \frac{B_j(\bar{n}_{j-1})}{(\alpha N)^2} \Bigg |\sum\limits_{m=0}^{(1-\alpha)N}\Bigg [ 
\sum\limits_{\bar{m}_j \in S_m^j} \bigg \{ A_j(n, \bar{n}_j) B_j(\bar{m}_{j-1})y^j(\bar{m}_j) A_j(m, \bar{m}_j) \bigg \} \\
&- \sum\limits_{\bar{p}_j \in S_n^j} \bigg ( B_j(\bar{p}_{j-1}) y^j(\bar{p}_j) A_j(n, \bar{p}_j) A_j(m, \bar{n}_j) \bigg )
\Bigg ] \Bigg | 
\end{align*}
where inequality a comes from Lemma \ref{denominator}, and equalities b and c come from applying equations \ref{eq:L}, \ref{eq:DL}, \ref{eq:Un}, and \ref{eq:DUn}. Equality d comes from factoring out $B_j(\bar{n}_{j-1})$ and observing that $\bar{n}_j \in S_n^j, S_m^j$ because we assumed it was in $[0,\alpha N]^j$, which is a subset of $S_k$ for every $k$. 

We observe that when $\bar{m}_j, \bar{p}_j \in [0,\alpha N]^j$, the expression above in the braces $\{\}$ will be equal to the expression in the parentheses $()$ by Lemma \ref{AB Same} and equation \ref{eq:marginals}. Additionally, we bound the expression in the braces $\{\}$. Using Lemmas \ref{outcome} and \ref{AB Bound}, we have that for any $\bar{m}_j$ 
\[A_j(n, \bar{n}_j) B_j(\bar{m}_{j-1})y^j(\bar{m}_j) A_j(m, \bar{m}_j) \leq \frac{1}{(\alpha N)^j},\]
with the same bound being true for the expression in the parentheses $()$. Thus, we see 
\begin{align*}
\left|\frac{\partial f(\y)}{\partial y_j(\bar{n}_j)} \right | 
\leq & \frac{B_j(\bar{n}_{j-1})}{(\alpha N)^2} \sum\limits_{m=0}^{(1-\alpha)N}\Bigg [ \max \Bigg (
\sum\limits_{\bar{m}_j \in S_m^j, \bar{m}_j \not \in [0,\alpha N]^j}  A_j(n, \bar{n}_j) B_j(\bar{m}_{j-1})y^j(\bar{m}_j) A_j(m, \bar{m}_j),  \\
& \sum\limits_{\bar{p}_j \in S_n^j, \bar{p}_j \not \in [0,\alpha N]^j} B_j(\bar{p}_{j-1}) y^j(\bar{p}_j) A_j(n, \bar{p}_j) A_j(m, \bar{n}_j) \Bigg )
\Bigg ]  \\
\leqa&  \frac{B_j(\bar{n}_{j-1})}{(\alpha N)^2} \sum\limits_{m=0}^{(1-\alpha)N} \frac{N^j[(1-\alpha)^j - \alpha^j]}{(\alpha N)^j} \\
\leqb& \frac{1}{(\alpha N)^{j+1}} \sum\limits_{m=0}^{(1-\alpha)N} \frac{N^j[(1-\alpha)^j - \alpha^j]}{(\alpha N)^j} \\
=& \frac{(1-\alpha)N}{(\alpha N)^{j+1}} \(\frac{(1-\alpha)^j}{\alpha^j} - 1 \)
\end{align*}
Inequality b comes from Lemma \ref{AB Bound}. Inequality a comes from the bound above, and observing that there are at most $N^j(1-\alpha)^j$ vectors in $S_m^j$ or $S_n^j$ and exactly $N^j \alpha^j$ vectors in $[0,\alpha N]^j$.
\end{proof}

\subsection{Bounding the Gradient}

We are now able to prove a bound on the gradient of the function in equation \ref{eq:f_n} in terms of $d$ and $\alpha$.
\begin{theorem}[Bounded Gradient] 
\label{gradient}
For any $n \in [0, (1-\alpha)N]$,
\[\|\nabla f_n(\y)\|_1 \leq \frac{2d(1-\alpha)}{\alpha}\(\frac{(1-\alpha)^d}{\alpha^d} - 1 \).\]
In particular, when $G$ has maximum degree $\Delta$, $\|\nabla f_n(\y)\|_1 < 1$ so long as $\alpha > \alpha_\Delta$, where $\alpha_\Delta$ is defined in equation \ref{eq:bound-alpha}.
\end{theorem}
\begin{proof}
First, fix some $j \in [1,d]$. Notice that there are exactly $(\alpha N)^j$ different $\bar{n}_j \in [0,\alpha N]^j$. Further, there are $(1-\alpha)^j N^j$ different $\bar{n}_j \in [0,(1-\alpha) N]^j$, meaning there are $N^j[(1-\alpha)^j - \alpha^j]$ vectors in the latter not in the former.

Using theorems \ref{loose} and \ref{tighter}, we obtain 
\begin{align*}
\sum\limits_{\bar{n}_j \in [0,(1-\alpha)N]^j} \left | \frac{\partial f(\y)}{\partial y^j(\bar{n}_j)} \right | 
&\leq (\alpha N)^j \frac{(1-\alpha)N}{(\alpha N)^{j+1}} \(\frac{(1-\alpha)^j}{\alpha^j} - 1 \) + N^j((1-\alpha)^j - \alpha^j) \frac{(1-\alpha)N}{(\alpha N)^{j+1}}\\
&= \frac{2(1-\alpha)}{\alpha}\(\frac{(1-\alpha)^j}{\alpha^j} - 1 \).
\end{align*}
Using this, we can conclude
\begin{align*}
\|\nabla f_n(\y)\|_1 &= \sum\limits_{j=1}^d \sum\limits_{\bar{n}_j \in [0,(1-\alpha)N]^j} \left | \frac{\partial f(\y)}{\partial y^j(\bar{n}_j)} \right |\\
&\leq \sum\limits_{j=1}^d \frac{2(1-\alpha)}{\alpha}\(\frac{(1-\alpha)^j}{\alpha^j} - 1 \)\\
&\leq \sum\limits_{j=1}^d \frac{2(1-\alpha)}{\alpha}\(\frac{(1-\alpha)^d}{\alpha^d} - 1 \)\\
&= \frac{2d(1-\alpha)}{\alpha}\(\frac{(1-\alpha)^d}{\alpha^d} - 1 \).
\end{align*}
\end{proof}

Evaluated at $\Delta = 3$, we obtain $\alpha_3 \approx 0.488$. Take $\alpha = 0.5 - c/\Delta^\gamma$, with $c$ a constant not dependent on $\Delta$ and $\gamma > 0$. When $\gamma < 2$, we have
\[\lim_{\Delta \rightarrow \infty} \frac{2\Delta(0.5 + c/\Delta^\gamma)}{(0.5 - c/\Delta)^\gamma}\(\frac{(0.5 + c/\Delta^\gamma)^\Delta}{(0.5 - c/\Delta^\gamma)^\Delta} - 1 \) = \infty.\]
But, when $\gamma = 2$ and $c < \frac{1}{8}$, we have
\[\lim_{\Delta \rightarrow \infty} \frac{2\Delta(0.5 + c/\Delta^2)}{(0.5 - c/\Delta)^2}\(\frac{(0.5 + c/\Delta^2)^\Delta}{(0.5 - c/\Delta^2)^\Delta} - 1 \) = 8c < 1.\]
Thus, $\alpha_\Delta$ is of the form $0.5 - O(1/\Delta^2)$.


%% file: sections/algorithms.tex
\algblock{Input}{EndInput}
\algnotext{EndInput}
\algblock{Output}{EndOutput}
\algnotext{EndOutput}
\newcommand{\Desc}[2]{\State \makebox[6em][l]{#1}#2}

In this section, we design an algorithm to approximate $Z(G)$. We first demonstrate how to use the contraction proven in Section \ref{correlation decay} to design a recursive algorithm which approximates $x(G, v \la n, \beta)$ with error decaying geometrically in the recursion depth. With this, we will present the approximation algorithm and formalize the argument from Section \ref{definitions}.

\begin{algorithm}
\caption{ApproxProb}
\label{algo:prob}
\begin{algorithmic}[1]
  \Input
  \Desc{$G=(V,E)$}{Bounded degree Graph with at least 1 vertex}
  \Desc{$k$}{Remaining Computation Depth}
  \Desc{$v$}{Vertex in the graph $G$}
  \Desc{$\alpha$}{Value in $[0,0.5)$}
  \Desc{$n^*$}{Integer value to assign vertex $v$}
  \Desc{$N$}{Positive Integer}
  \Desc{$\beta$}{Set of constraints for the vertices $V$, with $\beta_v$ denoting the assigned value }
  \EndInput
  \Output
  \Desc{$\tilde{f}_n(\x)$}{Estimate of $f_n(\x)$, probability vertex $v$ is assigned $n^*$ in the Gibbs distribution}
  \EndOutput
  \State Let $b = \max\{\beta_{u_i}\}$, where $u_i$ is the set of constrained neighbors of $v$ 
  \If{$n^* > N - b$}  Return 0
  \ElsIf{$\exists$ constrained adjacent vertices $u_1, u_2$ with $\beta_{u_1} + \beta_{u_2} > N$}
   Return $0$
  \ElsIf{$k=0$} Return $\frac{1}{\min\{(1-\alpha)N+1, N-b+1\}}$
  \Else
    \State Let $d$ be the number of unconstrained neighbors $v$ has, and let $v_1, \ldots v_d$ be those neighbors
    \For{$j \in [1,d]$}
        \For{$n_1, \ldots n_j \in [0,(1-\alpha)N]$}
        \State Let $y^j(\bar{n}_j) = \operatorname{ApproxProb}(G \setminus v, k-1, v_j, \alpha, n_i,N, (\beta, v_i \la n_i, \forall i < j)) $
        \EndFor
    \EndFor
    \State Return $f_n(\y)$
  \EndIf
\end{algorithmic}
\end{algorithm}

\subsection{Approximating Probabilities}

In this section, we introduce Algorithm \ref{algo:prob}, ApproxProb. This algorithm recursively computes an approximation $y(G, v \la n, \beta)$ of $x(G, v \la n, \beta)$. 

\begin{lemma}
\label{probability approximation}
Take some arbitrary graph $G$ with degree bounded by $\Delta$, as well as some parameters $N, \alpha, \beta$ satisfying $\|\nabla f^v_n\|_1 = c < 1$,
where $f^v_n$ is the function from equation \ref{eq:f_n}, for each vertex $v$ in every subgraph of $G$ with any $\beta$ for the particular values of $N$ and $\alpha$. Pick any computation depth $k > 0$. 
Algorithm \ref{algo:prob} called with parameters $G, k, \alpha, N,$ and $ \beta$ will produce an approximation $y = y(G, v \la n^*, \beta)$ of $x = x(G, v \la n^*, \beta)$  satisfying
\[|y - x| \leq c^k.\]
Further, it will do so in time $N^{O(k\Delta)}$.
\end{lemma}
\begin{proof}
We will first need to establish that the estimated marginal probabilities $\y$ computed on line 19 form a Feasible Marginal Distribution. We will do so inductively, starting with $k=1$, as it is the first time we compute such a $\y$. When $k = 1$, $\y$ will form a Feasible Marginal Distribution by construction. The Feasible Probability Distribution requirements will be satisfied by the value returned on line 14, since it will be the uniform probability measure over all feasible values of $n^*$. Further, for any $j \in \{1, \ldots d\}$ and $\bar{n}_{j}, \bar{m}_j \in [0,\alpha N]^j$, $y^j(\bar{n}_j) = y^j(\bar{m}_j) = \frac{1}{(1-\alpha)N}$, satisfying the final requirement.

Proceeding by induction for $k > 1$, we have that the Feasible Probability Distribution requirements will be satisfied by Lemma \ref{closure f} and the inductive hypothesis. Additionally, observe that if $n,m \leq \alpha N$, changing $u \la n$ to $u \la m$ in a constraint set $\beta$ will not change the result of running ApproxProb. This is because $u$ assigned to either value cannot violate any constraints. Finally, notice that the vector $\y$ computed in a call to ApproxProb does not depend on $n^*$, and $f_n = f_m$ for any $n,m \le \alpha N$. Thus, the vector $\y$ computed on line 19 must satisfy equation \ref{eq:marginals}, so $\y$ is a Feasible Marginal Distribution. 

We will now show the main result via induction. For the base case, suppose that $k=1$. Let $\x$ be the true values of the variables $x(\bar{n}_i)$ according to the Gibbs distribution of their corresponding graphs. Let $\y$ be the values of the variables computed in the execution of the algorithm on line 19. By the mean value theorem, there must exists some $\z$, which is a linear interpolation of $\y$ and $\x$ such that
\[f_n^v(\x) - f_n^v(\y) = \nabla f^v_n(\z) \cdot (\x -\y).\]
By Lemma \ref{linear interpolations}, $\z$ must also be a Feasible Marginal Distribution. Further, every entry in $\x$ and $\y$ is non-negative and bounded by 1. Thus,
\[|f_n^v(\x) - f_n^v(\y)| \leq \|\nabla f^v_n(\z)\|_1 \|\x -\y\|_\infty \leq c.\]

We now induct over the values of $k$. Once again, let $\x$ be the true values of the variables $x(\bar{n}_i)$, and let $\y$ be the values computed on line 19 during the outermost layer of the recursion. Notice that any entries of $\y$ which are 0 will also be 0 in $\x$. This, combined with the inductive hypothesis, implies that $\|\x - \y\|_\infty \leq c^{k-1}$. Thus, using the same logic as above, we can use the mean value theorem to show that $|f^v_n(\x) - f^v_n(\y)\| \leq c^k$ as desired.

Additionally, we wish to analyze the runtime of Algorithm \ref{algo:prob}. Let $T(k)$ denote the runtime of the algorithm with the input $k$, and notice that $T(0) = O(1)$. For runs with higher values of $k$, the limiting time step will be computing the recursive calls on line 20. There will be at most $(1-\alpha)^\Delta N^\Delta$ iterations, and each will take $O(\Delta^2)$ time to update the values of $\beta$. Therefore, we get that $T(k) = O((1-\alpha)^\Delta N^\Delta \Delta^2 T(k-1))$ implying that $T(k) = O\(\((1-\alpha)^\Delta N^\Delta \Delta^2\)^k\) $. Since we are interested in the runtime in terms of $\Delta$, $k$ and $N$, this gets us that $T(k) = N^{O(k\Delta)}$. 
\end{proof}

\begin{algorithm}
\caption{ApproxZ}
\label{algo:count}
\begin{algorithmic}[1]
  \Input
  \Desc{$G=(V,E)$}{Bounded degree Graph with at least 1 vertex}
  \Desc{$k$}{Computation Depth}
  \Desc{$\alpha$}{Value in $[0,0.5)$}
  \Desc{$N$}{Positive Integer}
  \EndInput
  \Output
  \Desc{$\tilde{Z}(G)$}{Estimate of $Z(G)$, the number of restricted relaxed independent sets in $G$}
  \EndOutput
  \State Let $z\_inv = 1$
  \State Let $\beta = ()$
  \For{$v \in V$}
    \State $z\_inv \gets z\_inv * \operatorname{ApproxProb}(G, k, v, \alpha, 0, N, \beta )$
    \State $\beta \gets (\beta, v \la 0)$
  \EndFor
  \State Return $1/z\_inv$
\end{algorithmic}
\end{algorithm}

\subsection{Counting Algorithm}

We now introduce Algorithm $\ref{algo:count}$ ApproxZ, which will approximately compute $Z(G)$, the number of restricted relaxed independent sets in the graph $G$. This algorithm directly computes the telescoping definition of $Z(G)$ using the approximated marginal probabilities produced by Algorithm \ref{algo:prob}. We will show the following main result.

\begin{theorem}
\label{counting algorithm}
For any graph $G$ of maximum degree $\Delta$ with $m$ vertices, Algorithm \ref{algo:count} will produce an estimate $\hat Z(G)$ satisfying
\[1 - O(1/m) \leq \frac{\hat Z(G)}{Z(G)} \leq O(1/m),\]
for any $N$ and any $\alpha > \alpha_\Delta$ for some $k$ which is $O(\log mN)$. Further, the algorithm will run in $N^{O(\Delta \log Nm)}$ time.
\end{theorem}

\begin{proof}
Since $\alpha > \alpha_\Delta$, we have that $\|\nabla f_n\|_1 \leq c < 1$ with respect to every vertex $v$ in ever subgraph of $G$.
Observe that equations \ref{eq:1 sum} and \ref{eq:non-inc} together imply that $y(G, v \la 0, \beta) \geq \frac{1}{(1-\alpha) N + 1}$. Thus, letting $k = \log_c\(\frac{1}{((1-\alpha)\alpha N + 1)m^2}\)$ in the execution of Algorithm \ref{algo:count}, Lemma \ref{probability approximation} says the estimates $y(G, v_k \la 0, \beta_{k-1})$ of $x(G, v_k \la 0, \beta_{k-1})$ will satisfy
\[|x(G, v_k \la 0, \beta_{k-1})-y(G, v_k \la 0, \beta_{k-1})| 
\leq \frac{1}{((1-\alpha) N + 1)m^2} 
\leq \frac{y(G, v_k \la 0, \beta_{k-1})}{m^2}.\]
Dividing both sides of the equation by $y(G, v_k \la 0, \beta_{k-1})$ results in
\[|\frac{x(G, v_k \la 0, \beta_{k-1})}{y(G, v_k \la 0, \beta_{k-1})} - 1| \leq  \frac{1}{m^2}.\]
As discussed in Section \ref{definitions}, this bound is sufficient to show that the estimate $\hat Z(G)$ satisfies
\[1 - O(1/m) \leq \frac{\hat Z(G)}{Z(G)} \leq O(1/m).\]
We finish by noting that since $c < 1$, $k$ is $O(\log mN)$ for constant $\alpha$. The bound on the running time follows directly from the bound in Lemma \ref{probability approximation}.
\end{proof}

%% file: sections/volumes.tex
\DeclarePairedDelimiter\floor{\lfloor}{\rfloor}
\newcommand{\ceil}[1]{#1}
\newcommand{\U}{\mathcal{U}}
\renewcommand{\L}{\mathcal{L}}
\renewcommand{\Q}{\mathcal{Q}}

Now that we have an algorithm to count restricted relaxed independent sets, we move to showing how one can approximate the volume of $\P_{\rm IS}(G, \alpha)$ using this discrete approximation.

Fix an arbitrary graph $G = (V,E)$ with $V = \{v_1, \ldots v_m\}$, where each vertex has degree at most $\Delta$. We define a set of graphs $G_n = (\{v_1, \ldots v_n\}, E_n)$ where $E_n = \{(v_i,v_j) \in E| i,j \leq n\}$, so that $G_m = G$. Additionally, fix $\alpha \geq \alpha_\Delta$ and $N$. Since we will be inducting over the size of the graph, we will redefine $\beta$ to be a set of constraints of the form $x_i \leq c$, and have $\beta_i$ denote the tightest constraint on $x_i$. In particular, meaning any $\x$ satisfying every constraint in $\beta$ satisfies $x_i \leq \beta_i$. Similarly, we will redefine $v_i \la c$ to be the set of constraints $\{x_j \le N-c|(v_i, v_j) \in E\}$. We assume that $\alpha N \leq \beta_i \leq \ceil{(1-\alpha)N}$. Under this setup, we define the following family of polytopes,
\[\Q(G_n, \alpha, \beta) = \{\x \in \R^n: 0 \leq x_i \leq \beta_i\  \forall i\in [1,n], x_i + x_j \leq N \  \forall (v_i,v_j) \in E_n\}.\]
We will show that for any such polytope, we can approximate its volume by counting the number of restricted relaxed independent sets in $G$. To do this, we define the following two sets,
\begin{align*}
\U(G_n, \alpha, \beta) &= \{\x \in \Z^n: 0 \leq x_i \leq \beta_i \  \forall i\in [1,n], x_i + x_j \leq N \  \forall (v_i,v_j) \in E_n\},\\
\L(G, \alpha, \beta) &= \{\x \in \Z^n: 0 \leq x_i \leq \beta_i - 1 \ \forall i\in [1,n], x_i + x_j \leq N-2 \  \forall (v_i,v_j) \in E_n\}.
\end{align*}
Notice that $\frac{\Q(G, \alpha, \beta)}{N^m} \subseteq \P_{\rm IS}(G, \alpha)$, and when  $\beta$ is such that $\beta_i = (1-\alpha)N, \forall i$, $|\U(G, \alpha, \beta)| = Z(G)$ and $\frac{\Q(G, \alpha, \beta)}{N^m} = \P_{\rm IS}(G, \alpha)$. 
\begin{lemma}[Volume Bound]
\label{volume bound}
For any graph $G$ and any   $n, \beta, \alpha, N$, 
\[|\L(G, \alpha, \beta)| \leq \vol(\Q(G_n, \alpha, \beta)) \leq |\U(G_n, \alpha, \beta)|.\]
\end{lemma}
\begin{proof}
We first show the lower bound. Take any vector $\x \in \L(G, \alpha, \beta)$. It follows immediately from construction that $\x \in \Q(G_n, \alpha, \beta)$. Further, For any vector $\y \in [0,1)^n$, we have that $x_i + y_i \leq \beta_i$ and $(x_i + y_i) + (x_j + y_j) \leq x_i + x_j + 2 \leq N$ for every $(i,j) \in E_n$. Thus, $\x + \y \in \Q(G_n, \alpha, \beta)$ for every such $\y$. Since we can associate a distinct unit cube within $\Q(G_n, \alpha, \beta)$ to each point $\x \in \L(G, \alpha, \beta)$, we get that $|\L(G, \alpha, \beta)| \leq \vol(\Q(G_n, \alpha, \beta))$ as desired.

Next we show the upper bound. Consider any vector $\y \in \Q(G_n, \alpha, \beta)$, and let $\floor{\y}$ be the vector $\y$ with each entry truncated to the highest  integer value from below. We clearly have  $\floor{\y} \in \U(G_n, \alpha, \beta)$. Assigning a cube of side length one  to each $\x \in \U(G_n, \alpha, \beta)$, we obtain a space which contains $\Q(G_n, \alpha, \beta)$. Thus, $\vol(\Q(G_n, \alpha, \beta)) \leq |\U(G_n, \alpha, \beta)|$.
\end{proof}

Now, we move on to the main theorem of this section.
\begin{theorem}[Volume Approximation]
\label{volume approximation}
For any graph $G$ and any  of $n, \beta, \alpha, N$,
\[
\left(1 - \frac{n}{\ceil{\alpha N}}\right) |\U(G_n, \alpha, \beta)| \leq \vol(\Q(G_n, \alpha, \beta)) \leq |\U(G_n, \alpha, \beta)|.\]
\end{theorem}
\begin{proof}
Let $\x \in \U(G_n, \alpha, \beta)$ be picked uniformly at random. Further, let $\mu_\beta^n = \Pr(\x \not \in \L(G, \alpha, \beta)) = \frac{|\U(G_n, \alpha, \beta)| - |\L(G, \alpha, \beta)|}{|\U(G_n, \alpha, \beta)|}$. We will show $\mu_\beta^n \leq \frac{n}{\ceil{\alpha N}}$ by induction on $n$. This will directly imply the result by Lemma \ref{volume bound} since $|\L(G_n, \alpha, \beta)| = (1 - \mu_\beta^n) |\U(G_n, \alpha, \beta)|$. In the base case, if $n = 1$, then $\mu_\beta^1 = \Pr(x_1 \la \beta_1) = \frac{1}{\beta_1+1} \leq \frac{1}{\ceil{\alpha N}}$, since $x_1$ is uniformly distributed over $[0,\beta_1]$ and $\beta_1 \geq \alpha N$. 

We now proceed by induction. First, notice that for any $\x \in \U(G_n, \alpha, \beta)$, replacing the $n$th entry of $\x$ with any smaller value will produce a new $\x' \in \U(G_n, \alpha, \beta)$. In particular, this means for any $0 \leq j \leq i$, it must be that $\Pr(x_n \la i) \leq \Pr(x_n \la j)$. Thus, 
\[\Pr(x_n \la i) \leq \frac{\Pr(x_n \la i)}{\sum_{j=0}^i\Pr(x_n \la j)} \leq \frac{\Pr(x_n \la i)}{(i+1)\Pr(x_n \la i)} = \frac{1}{i+1},\]
and since $\beta_n \geq \alpha N$, we get that $\Pr(x_n \la \beta_i) \leq \frac{1}{\ceil{\alpha N}}$. 

Next, consider any $\x \in \U(G_n, \alpha, \beta)$ with $x_n = k$ where $\x \not \in \L(G, \alpha, \beta)$. It must be that either $x_n = \beta_n$ or $(x_1, \ldots x_{n-1}) \not \in \L(G_{n-1}, \alpha, (\beta, v_n \la k))$. Notice that $k \leq \ceil{(1-\alpha)N}$, since $\beta_n \leq \ceil{(1-\alpha)N}$ by assumption. Therefore, we can use the inductive hypothesis to see
\begin{align*}
\mu_\beta^n &= \Pr(x_n \la  \beta_n) + \sum\limits_{i=0}^{\beta_i-1} \Pr(x_n \la i) \mu^{n-1}_{(\beta, x_n \la i)} \\
&\leq \frac{1}{\alpha N} + \sum\limits_{i=0}^{\beta_i-1} \Pr(x_n \la i) \frac{n-1}{\alpha N}\\
&\leq  \frac{n}{\alpha N}.
\end{align*}
\end{proof}

With this analysis, we are ready to complete the proof of our main result, Theorem \ref{main result}. 

\begin{proof}[Proof of Theorem \ref{main result}]
We will actually prove an asymptotically stronger result, namely that there is an algorithm running in $m^{O(\log m)}$ time which produces a value $\tilde V$ satisfying
\[1 - O(1/m) \le \frac{\tilde V}{\vol(\P_{\rm IS}(G,\alpha))} \leq 1 + O(1/m)\]
when $G$ is an $m$ node graph with a maximum degree of $\Delta$ and $\alpha > \alpha_\Delta$. The proof which follows can be modified to a proof of the main theorem by setting $N$ to be $O(m/\epsilon)$ and modifying Theorem $\ref{counting algorithm}$ to have $k$ be $O(\log m/\epsilon)$. 

We start by setting $N = \frac{m^2}{\alpha}$. Let $\beta$ be such that $\beta_i = \ceil{(1-\alpha)N}$ for every $i$. We have that $\P_{\text{IS}}(G,\alpha) = \frac{\Q(G, \alpha, \beta)}{N^m}$. By Theorem \ref{volume approximation}, we have 
\[\(1 - \frac{m}{\alpha N}\)|\U(G, \alpha, \beta)| = \(1 - \frac{1}{m}\) |\U(G, \alpha, \beta)| \leq \vol(\Q(G, \alpha, \beta)) \leq |\U(G, \alpha, \beta)|,\]
as a result
\[|\U(G, \alpha, \beta)| - \vol(\Q(G, \alpha, \beta))| = |\U(G, \alpha, \beta)| - \vol(\Q(G, \alpha, \beta)) \leq \frac{1}{m} |\U(G, \alpha, \beta)| \leq \frac{2}{m} \vol(\Q(G, \alpha, \beta)), \]
where the last inequality follows from $1 - \frac{1}{m} \geq \frac{1}{2}$ (unless the graph is trivial) implying that $2\vol(\Q(G, \alpha, \beta)) \geq |\U(G, \alpha, \beta)|$. Additionally, we have that $|\U(G, \alpha, \beta)| = Z(G)$ by construction.  Theorem \ref{counting algorithm} implies the existence of a deterministic algorithm to produce an estimate $\tilde{Z}(G)$ satisfying $|\tilde{Z}(G) - |\U(G, \alpha, \beta)|| \leq  c|\U(G, \alpha, \beta)|/m$ for some universal constant $c$. Thus, we get
\begin{align*}
|\vol(\Q(G, \alpha, \beta)) - \tilde{Z}(G, \alpha)| &\leq |\vol(\Q(G, \alpha, \beta)) - |\U(G, \alpha, \beta)|| + ||\U(G, \alpha, \beta)| - \tilde{Z}(G,\alpha)|\\
&\leq \frac{2}{m} \vol(\Q(G, \alpha, \beta)) + \frac{c}{m} |\U(G, \alpha, \beta)|\\
&\leq \frac{2}{m} \vol(\Q(G, \alpha, \beta)) + \frac{2c}{m} \vol(\Q(G, \alpha, \beta))\\
&= \frac{2c +2}{m} \vol(\Q(G, \alpha, \beta)),
\end{align*}
where the last inequality again holds because $2\vol(\Q(G, \alpha, \beta)) \geq |\U(G, \alpha, \beta)|$. Therefore, we can take $\tilde V = \frac{\tilde{Z}(G, \alpha)}{N^m}$ to be our estimate of the volume of $\P_{\text{IS}}$ to see
\[|\vol(\P_{\text{IS}}) - \frac{\tilde{Z}(G, \alpha)}{N^m}| = \frac{1}{N^m} |\vol(\Q(G, \alpha, \beta)) - \tilde{Z}(G, \alpha)| \leq \frac{2c+2}{mN^m } \vol(\Q(G, \alpha, \beta)) = \frac{2c+2}{m} \vol(\P_\text{IS}). \]

Additionally, substituting in the value of $N$, the algorithm from Theorem \ref{counting algorithm} will run in $m^{O(\Delta \log m)}$ time.

\end{proof}

%% file: sections/obstacles.tex


In this section, we discuss some challenges we encounter when designing an algorithm to compute the volume of the unrestricted ($\alpha = 0$) independent set polytope.
While the gradient bound given in Section \ref{correlation decay} is loose and some adjustment to the argument may admit an asymptotic improvement on our values of $\alpha_\Delta$, it clearly will not generalize to $\alpha = 0$. In fact, we have been able to show computationally that the values of $|\nabla f_n|_1$, with $f_n$ defined as in \ref{eq:f_n} exceeds 1 in bounded degree graphs when $\alpha = 0$. We set $N = 64$ and take a 4-regular tree of depth 50, and assigned each leaf vertex 0 with probability 1. We were than able to repeatedly apply equation \ref{eq:f_n} to produce a probability distribution for the assignment of the vertices at each level of the tree. We know that correlation decay holds on trees \cite{gamarnik2019uniqueness}, so the probability distributions quickly converges to a fixed point distribution. At depth 50, the probability of a vertex being assigned a value $n$ had converged within 5 significant decimal places. 

We evaluated $|\nabla f_n(\x)|_1$ for each $n$ with this probability distribution $\x$. We found that for $n \leq 13$, $|\nabla f_n(\x)|_1 > 1$. When we modified $N$, we found that roughly the same proportion of values $n$ satisfied $|\nabla f_n(\x)|_1 > 1$. Increasing the degree of the tree increased the number of $n$ where $|\nabla f_n(\x)|_1 > 1$. 

Since we initialized each leaf node with the same value, every vertex at a specified depth in our tree will have the same probability distribution over $[0,N]$. Thus, we can modify equation \ref{eq:f_n} to be only a function of $x_0, \ldots x_N$, where $x_i$ is the probability a vertex of depth one below our current vertex is assigned $i$. This allows us to construct the $N+1 \times N+1$ matrix $M$, where $M_{ij} = \frac{\partial f_i(\x)}{\partial x_j}$. In all of the cases above, we found that this matrix $M$ had a largest singular value greater than 1, and thus had an operator norm exceeding 1. This seems to suggest that  methods of showing correlation decay which relies on showing a universal single step contraction, such as one similar to (\cite{chen2023strong}) does not succeed.
It is possible that methods based on multi-step contraction or based on other single step but non-spectral type of contraction do succeed. We leave it as a very interesting question for further research.
